\documentstyle[12pt,epsfig,rotating]{article}

\begin{document}
\begin{flushright}
University Siegen\\
Preprint SI 96-12
\end{flushright}

\vspace*{3cm}
\begin{center}
{\Large {\bf HOW TO MEASURE DIFFRACTION}}\\[0,3cm]
{\Large {\bf IN TWO-PHOTON COLLISIONS AT LEP}}\\[0,5cm]

\vspace*{3cm}

{\bf R.\ Engel\footnote{e-mail: eng@tph200.physik.uni-leipzig.de}}\\
{\em Universit\"at Leipzig, Fachbereich Physik, D-04109 Leipzig, Germany
\\ and Universit\"at Siegen, Fachbereich Physik, D-57068 Siegen, Germany}

\vspace*{0.5cm}
{\em and}
\vspace*{0.5cm}

{\bf A.\ Rostovtsev\footnote{e-mail: rostov@dice2.desy.de}}\\
{\em LPNHE University-VI, 75252 Paris, France\\
and Institute of Theoretical and Experimental Physics,\\
        117259 Moscow, Russia}

\end{center}

\vspace*{1.5cm}

\begin{abstract}
\noindent
The possibility to
measure diffraction dissociation in collisions of real and virtual 
photons at LEP2 is pointed out. 
\end{abstract}

\clearpage

\noindent
  At the energy of $e^+e^-$ collisions well above $Z^0$ resonance mass 
the dominant mechanism of hadron production is the interaction of two
photons radiated off the lepton beams. 
  The photons available for the collision can be selected within a
broad range of the energy and virtuality.  
  This makes the photon collisions an unique testing ground for
investigations of strong interaction phenomena.
  The high energy interaction of photons with low virtualities 
represents the bulk of the events and is understood in terms
of the Regge formalism. Similar to hadron interactions, a rise
of the total two-photon cross section with the collision 
energy and a large diffractive contribution are expected. 
  The recent HERA data on the $ep$ scattering show that these phenomena
also persist for highly virtual photon interaction, where the photon
probes the short distance structure of the interaction
and perturbative QCD calculations can be applied.
  In the present letter we point out the possibility to measure
photon diffraction dissociation in high energy two-photon
collisions.
 \begin{figure}[!hbt]  \centering
 \boldmath
\epsfig
{file=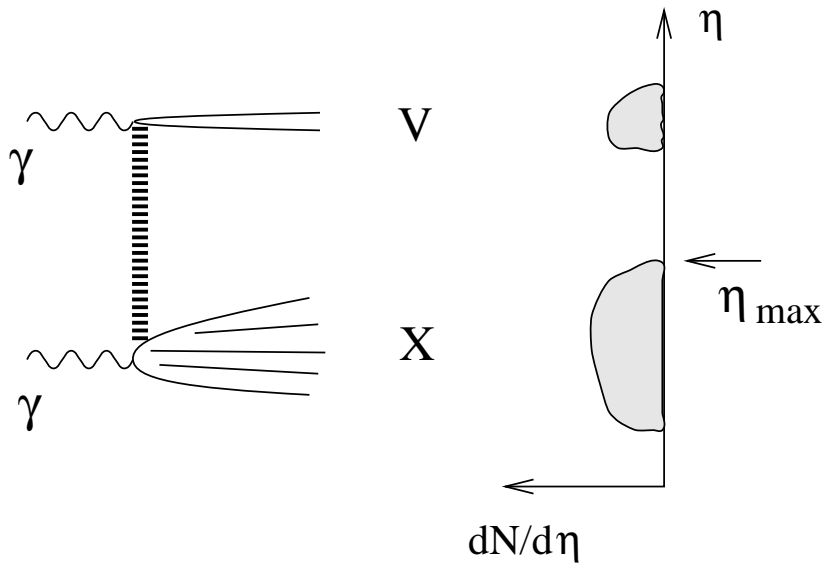,width=110mm}
 \normalsize \unboldmath
\caption
  {\small Photon single diffraction dissociation and the expected
pseudorapidity distribution of final state hadrons.}
\label{gg-etamax}
 \end{figure}

 Diffractive reactions are investigated since long time in 
hadron-hadron scattering, photoproduction and, recently, also in
deep-inelastic scattering. 
The $e^+e^-$ collider LEP2 and planned linear colliders provide an
ideal tool to investigate these processes also in photon-photon collisions. 
Diffractive processes have typically large partial cross sections
and manifest themselves in spectacular topologies with rapidity gaps.
In Fig.~\ref{gg-etamax}, an example of single photon diffraction
dissociation is shown.

The expected cross section for 
photon diffraction dissociation can be estimated from simple 
factorization arguments. In Regge phenomenology, the
relative contribution $R$ of single-diffractive dissociation (SD) of photons 
to the total cross section
can be approximated by
\begin{equation}
R \approx 
\frac{\sigma_{p,SD}(\gamma p \rightarrow V X)
}{\sigma_{tot}(\gamma p \rightarrow X)}
\approx
\frac{\sigma_{\gamma,SD}(\gamma \gamma \rightarrow V X) 
}{\sigma_{tot}(\gamma \gamma \rightarrow X)},
\end{equation}
where $\sigma_{p,SD}$ and $\sigma_{\gamma,SD}$ are the cross sections for 
diffraction dissociation of protons and photons, respectively.
$V$ denotes the low-mass vector mesons. 
Using the recent H1 measurement of the partial photoproduction cross 
sections~\cite{Aid95b}, one gets $\sigma_{\gamma,SD} \approx 0.1\cdot 
\sigma_{tot}$ in photon-photon collisions.
A  cross section 
of similar size is also expected for quasi-elastic vector meson production 
\cite{Engel95d,Schuler96a}.

Experimentally, events with diffraction dissociation can be identified 
using the rapidity gap technique. In non-diffractive 
reactions, rapidity gaps between the final state hadrons are 
exponentially suppressed. In contrast, the differential
cross section $d\sigma_{\gamma,SD}/d\eta_{gap}$ of diffraction 
dissociation is independent of the width $\eta_{gap}$ of such rapidity gaps.
Therefore, diffractive particle production can be measured triggering on 
large rapidity gaps. 
However, the limited angular detector acceptance of the main detector
makes the
measurement of the complete gap hardly possible. 
As shown by the HERA Collaborations~\cite{Derrick93a,Ahmed94a}, 
the measurement of the 
 so-called $\eta_{max}$ distribution can be used to obtain experimental 
evidence for diffraction.  
 The variable $\eta_{max}$ is defined as the maximum pseudorapidity of the
final state hadrons produced in the two-photon collision excluding the
decay products of the quasi-elastically scattered vector meson $V$
(see Fig.~\ref{gg-etamax}). 
Note that it is not needed to measure the hadrons in
the entire pseudorapidity range allowed by phase space. The
combination of a forward detector with the central main detector parts is
well suited to find evidence for diffraction (it is also unimportant whether 
there is a gap in the pseudorapidity coverage between these detector parts).
In other words, the variable $\eta_{max}$ used here
measures the pseudorapidity edge of the multi-hadronic system produced
in central main detector.

%
 \begin{figure}[htb]  \centering
 \boldmath
\unitlength1mm
 \begin{picture}(130,90)(5,5)
\put( 7,45){\begin{sideways}
 {\large $d\sigma_{ee}\!/d\eta_{max}$~~~~(nb)}
\end{sideways}}
\put( 115,6){\large $\eta_{max}$}
\put(15,12){\epsfig{file=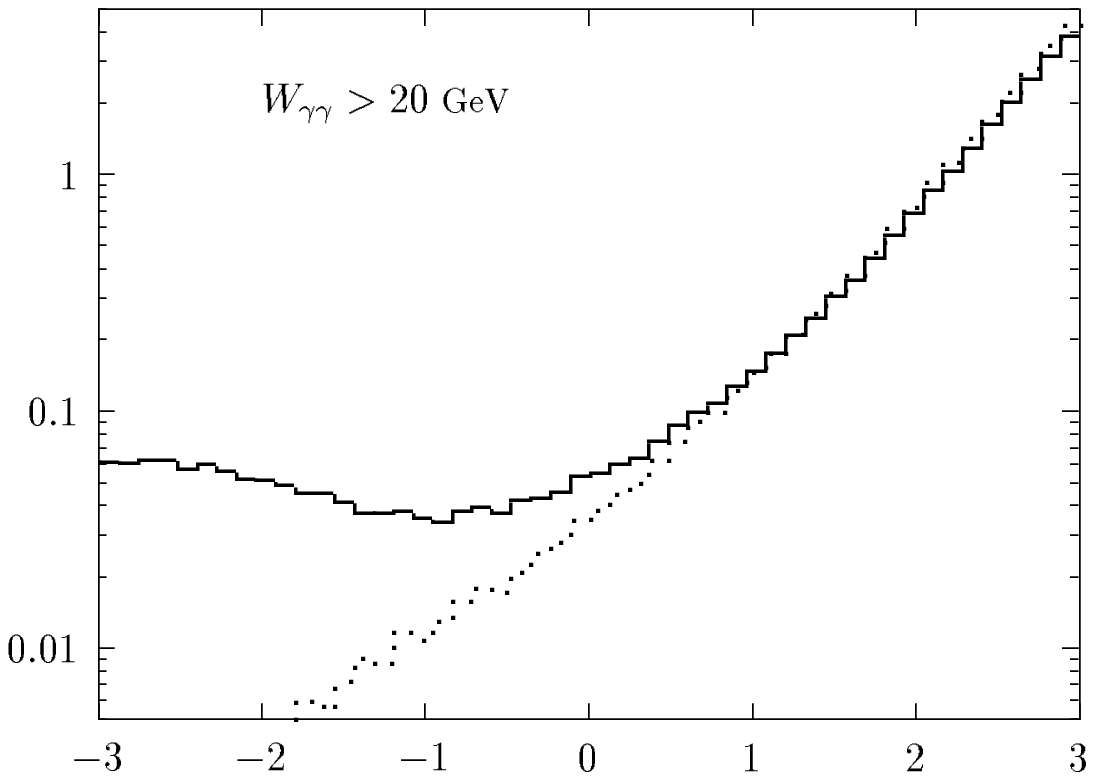,width=110mm}}
\end{picture}
 \normalsize \unboldmath
\caption{
\label{etamax1}
\small
The $\eta_{max}$ cross section as calculated using the {\sc Phojet}
MC event generator in
two photon collisions with $W \ge 20$~GeV (full curve). The dotted
curve shows the results of calculations for
non-diffractive $\gamma\gamma$ interactions.
}
 \end{figure}
%
The prediction of the $\eta_{max}$ cross section is shown in
Fig.~\ref{etamax1}
for photon-photon interactions in $ee$ collisions at LEP2 
($\sqrt{s}_{ee} = 175$ GeV) with 
an invariant $\gamma\gamma$ mass $W\ge 20$~GeV. The calculations 
were made using the {\sc Phojet} Monte Carlo event 
generator~\cite{Engel95a,Engel95d}.  
The $\eta_{max}$ distribution is obtained using the hadrons 
produced at 
pseudorapidities in the central range $-3 \le \eta \le 3$.
Only events having also hadrons produced in  
very forward direction ($3.5 \le \eta \le 4$) are considered. These 
particles can be measured with a small angle tagging calorimeter as 
installed at the LEP detectors~\cite{Buskolic95a,Abreu96a,Adeva90a,Ahmet91a}.
In diffractive events with a large rapidity gap, 
the particles produced in the 
very forward region are decay products of diffractively 
produced vector mesons. 
Note that it is not necessary to reconstruct these vector mesons.
The exponential suppression of the rapidity gap in non-diffractive events 
is clearly seen (dotted curve). Almost all events with $\eta_{max} < 0$ 
belong to diffraction (diffraction dissociation and quasi-elastic vector
meson production). 

%
 \begin{figure}[htb]  \centering
 \boldmath
\unitlength1mm
 \begin{picture}(130,90)(5,5)
\put( 7,45){\begin{sideways}
 {\large $d\sigma_{ee}\!/d\eta_{max}$~~~~(nb)}
\end{sideways}}
\put( 115,6){\large $\eta_{max}$}
\put(15,12){\epsfig{file=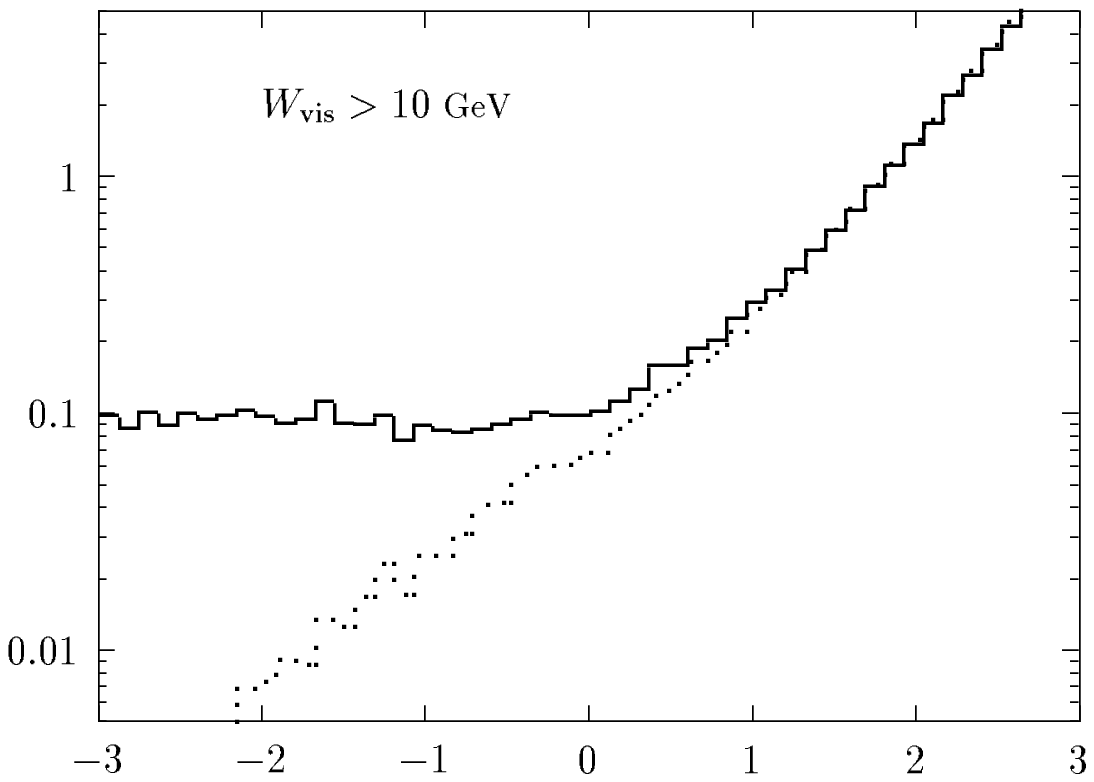,width=110mm}}
\end{picture}
 \normalsize \unboldmath
\caption{
\label{etamax-w10}
\small
The $\eta_{max}$ cross section as calculated using the {\sc Phojet}
MC event generator in
two photon collisions with $W_{\rm vis} \ge 10$~GeV (full curve). The dotted
curve shows the results of calculations for
non-diffractive $\gamma\gamma$ interactions.
}
 \end{figure}
%
It should be mentioned that a lower cut on the two-photon invariant mass
$W_{\gamma\gamma}$ is important to measure diffraction. It is beyond the
scope of this note to discuss the problem of the measurement of the
visible invariant mass $W_{\rm vis}$ and its relation to
$W_{\gamma\gamma}$. Clearly, the visible invariant mass depends on many
details of the experimental apparatus.
As a very simple example, we show in Fig.~\ref{etamax-w10} the $\eta_{max}$
cross section for two-photon events with $W_{\rm vis} > 10$ GeV. The
visible mass has been calculated from all charged hadrons
scattered at pseudorapidities $-3 \le \eta \le 3$ and $3.5
\le \eta \le 4$ assuming 100\% detection efficiency. 


Finally, we note that
the $\eta_{max}$ spectra in the Figs.~\ref{etamax1},~\ref{etamax-w10}
have been obtained
without using information on the kinematics of the scattered beam leptons
and, therefore, is dominated by the quasi-real photon interactions.
However, tagging one (or both) scattered leptons will allow to
measure also diffractive phenomena at small distances.
The investigation of transition from the real photon to highly
virtual photon initiated diffraction will provide important
information on the underlying physics of strong interactions.

{\bf Acknowledgements:}\\
One of the authors (RE) was supported by the Deutsche 
Forschungsgemeinschaft under contract No.~Schi 422/1-2.
We thank the DESY directorate for the hospitality.



\end{document}